\documentclass[a4paper,12pt,eqno]{article}
\usepackage{theorem}
\theoremheaderfont{\scshape}

\title{{\Large {\bf Temporal Fluctuations of Continuous-Time \\ Quantum Random Walks on Circles}
}}

\author{{\small Norio Inui, Koichiro Kasahara,$^{*}$ Yoshinao Konishi,$^{**}$ and Norio Konno$^{*}$} \\
{\small
{\it University of Hyogo, Yokohama National University,$^{*}$ Himeji Institute of Technology$^{**}$}}
}

\date{\empty }
\begin{document}
\maketitle

\par\noindent
\begin{small}
{\bf Abstract}. 
This work deals with both instantaneous uniform mixing property and temporal standard deviation for continuous-time quantum random walks on circles in order to study their fluctuations comparing with discrete-time quantum random walks, and continuous- and discrete-time classical random walks.

\footnote[0]{
{\it Key Words and Phrases.} 
Quantum walk; Temporal fluctuation; Circle.
}

\end{small}

\section{Introduction}

In this letter we consider mainly temporal fluctuations of continuous-time quantum random walks (QRWs). Farhi and Gutmann \cite{FG} proposed the model of a continuous-time QRW which might be useful for quantum computation. Recently Childs et al. \cite{CDC} showed that an exponential speedup was achieved by using continuous-time QRWs. Moore and Russell \cite{MR} investigated the mixing times and distributions for the continuous-time case on the hypercube first. Furthermore, Ahmadi et al. \cite{ABT} and Adamczak et al. \cite{AAH} studied mixing properties on various graphs, e.g., complete graph, $n$-cube, cycle. Gerhardt and Watrous \cite{GW} proved non-instantaneous uniform mixing on some Cayley graphs of the symmetric groups $S_n$ for $n \ge 3$. On the other hand, a discrete-time model was introduced by Ambainis et al. \cite{ABN} and Aharonov et al. \cite{AAK}, and has been explored by several groups (for examples, \cite{Ko,GJS,Ke,TFM,BGK,RSS}).

The present letter treats a continuous-time QRW on $C_N$ which is the cycle with $N$ vertices, i.e., $C_N =\{ 0, 1, \ldots , N-1 \}.$ Let $A$ be the $N \times N$ adjacency matrix of $C_N$. The continuous-time QRW considered here is given by the following unitary matrix:
\begin{eqnarray*}
U(t) = e^{-itA/2}
\label{eqn:unitary}
\end{eqnarray*}
Following the paper of Ahmadi et al. \cite{ABT}, we take $-itA/2$ instead of $itA/2$ . The amplitude wave function at time $t$, $| \Psi_N (t) \rangle$, is defined by
\begin{eqnarray*}
| \Psi_N (t) \rangle = U(t) | \Psi_N (0) \rangle 
\label{eqn:evolution}
\end{eqnarray*}
The $(n+1)$-th coordinate of $| \Psi_N (t) \rangle$ is denoted by $| \Psi_N (n,t) \rangle$ which is the amplitude wave function at vertex $n$ at time $t$ for $n=0,1, \ldots , N-1.$ In this letter we take $| \Psi_N (0) \rangle = [1,0,0, \ldots ,0]^{T}$ as an initial state, where $T$ denotes the transposed operator. The probability that the particle is at vertex $n$ at time $t$, $P_N (n,t)$, is
 given by
\begin{eqnarray*}
P_N (n,t) = \langle \Psi_N (n,t) | \Psi_N (n,t) \rangle 
\label{eqn:evolution}
\end{eqnarray*}
For the details of the definition for continuous-time QRWs are given in  \cite{ABT,GW}.

Here we give a connection between a continuum time limit model for discrete-time QRW on ${\bf Z}$ (= the set of integers) given by Romanelli et al. \cite{RSS} and continuous-time QRW on circles. Ahmadi et al. \cite{ABT} showed
\begin{eqnarray}
| \Psi_N (n,t) \rangle 
= {1 \over N} \sum_{j=0} ^{N-1} 
e^{-i(t \cos \xi_j - n \xi_j)},
\label{eqn:daiji}
\end{eqnarray}
for any $t \ge 0$ and $n=0,1, \ldots, N-1$, where $\xi_j =2 \pi j/N.$ As Ahmadi et al. \cite{ABT} pointed out, $| \Psi_N (n,t) \rangle$ can be also expressed by using the Bessel function, that is,
\begin{eqnarray}
| \Psi_N (n,t) \rangle =  \sum_{k=n \> (mod N)} (-i)^k J_k (t) = {1 \over 2} \sum_{k= \pm n \> (mod N)} (-i)^k J_k (t),
\label{eqn:contiquantumamp}
\end{eqnarray}
where $J_{\nu}(z)$ is the Bessel function (see  \cite{AAR}). In fact, combining the generating function of the Bessel function:
\begin{eqnarray*}
\exp \left[ {t \over 2} \left(z - {1 \over z} \right) \right] =  
\sum_{\nu \in {\bf Z}} z^{\nu} J_{\nu} (t),
\label{eqn:gf}
\end{eqnarray*}
and $J_{\nu} (-z) = (-1)^{\nu} J_{\nu} (z)$ gives
\begin{eqnarray*}
e^{-it \cos \xi_j} =  
\sum_{\nu \in {\bf Z}} (-1)^{\nu} e^{i \nu \xi_j} J_{\nu} (t).
\label{eqn:toyuu}
\end{eqnarray*}
By using Eq. (\ref{eqn:daiji}), the amplitude wave function is then calculated as 
\begin{eqnarray}
| \Psi_N (n,t) \rangle =  {1 \over N} \sum_{j=0} ^{N-1} \sum_{\nu \in {\bf Z}} (-i)^{\nu}  e^{i (n + \nu)  \xi_j} J_{\nu} (t).
\label{eqn:mousukosidesu}
\end{eqnarray}
In a similar way, we have
\begin{eqnarray}
| \Psi_N (n,t) \rangle =  {1 \over N} \sum_{j=0} ^{N-1} \sum_{\nu \in {\bf Z}} (-i)^{\nu}  e^{i (n - \nu)  \xi_j} J_{\nu} (t).
\label{eqn:mousukosidesuyo}
\end{eqnarray}
From Eqs. (\ref{eqn:mousukosidesu}) and (\ref{eqn:mousukosidesuyo}), we obtain 
Eq. (\ref{eqn:contiquantumamp}). Moreover, Eq. (\ref{eqn:contiquantumamp}) gives
\begin{eqnarray}
P_{N}(n,t)
&=& \sum_{j,k \in {\bf Z}} 
\cos \left( {\pi \over 2} (j-k)N \right) J_{jN+n}(t) J_{kN+n}(t)
\nonumber \\
&=&
\sum_{k \in {\bf Z}} J_{kN+n}(t)^2
+
\sum_{j \not=k : j,k \in {\bf Z}} 
\cos \left( {\pi \over 2} (j-k)N \right) J_{jN+n}(t) J_{kN+n}(t).
\label{eqn:contiquantumprobbessel}
\end{eqnarray}
The following formula is well known (see page 213 in Ref. \cite{AAR}):
\begin{eqnarray}
\sum_{k \in {\bf Z}} J_{k}(t)^2 = 1,
\label{eqn:besselprob}
\end{eqnarray}
for $t \ge 0$. That is, $\{J_{k}(t)^2: k \in {\bf Z} \}$ is a probability distribution on ${\bf Z}$ for any time $t$. Eq. (\ref{eqn:besselprob}) can be rewritten as 
\begin{eqnarray}
\sum_{n=0} ^{N-1} \sum_{k \in {\bf Z}} J_{kN+n}(t)^2 = 1,
\label{eqn:besselgeneral}
\end{eqnarray}
for $t \ge 0$. Eqs. (\ref{eqn:contiquantumprobbessel}) and (\ref{eqn:besselgeneral}) give
\begin{eqnarray*}
\sum_{n=0} ^{N-1} \sum_{j \not=k : j,k \in {\bf Z}} 
\cos \left( {\pi \over 2} (j-k)N \right) J_{jN+n}(t) J_{kN+n}(t) = 0.
\end{eqnarray*}

Very recently, Romanelli et al. \cite{RSS} studied a continuum time limit for a discrete-time QRW on ${\bf Z}$ and obtained the position probability distribution. When the initial condition is given by $\tilde{a}_l (0) = \delta_{l,0}, \tilde{b}_l (0) \equiv 0$ in their notation for the Hadamard walk, the distribution becomes the following in our notation:
\begin{eqnarray*}
P (n,t) = J_{n}(t/\sqrt{2})^2.
\label{eqn:romanelli}
\end{eqnarray*}
More generally, we consider a discrete-time QRW whose coin flip transformation is given by the following unitary matrix:
\begin{eqnarray*}
U=
\left[
\begin{array}{cc}
a & b \\
c & d
\end{array}
\right]
\end{eqnarray*}
\par\noindent
Note that $a=b=c=-d=1/ \sqrt{2}$ case is equivalent to the Hadamard walk. In a similar fashion, we have
\begin{eqnarray}
P (n,t) = J_{n}(at)^2.
\label{eqn:romanelligen}
\end{eqnarray}
In fact, Eq. (\ref{eqn:besselprob}) guarantees $\sum_{n \in {\bf Z}} P(n,t) = 1$ for any $t \ge 0$. From the probability theoretical point of view, it is interesting that a continuum time limit model for a discrete-time QRW on ${\bf Z}$ gives a simple model having a squared Bessel distribution in the above meaning. For large $N$, Eq. (\ref{eqn:contiquantumprobbessel}) is similar  to Eq. (\ref{eqn:romanelligen}) except for the second term of the right-hand side of Eq. (\ref{eqn:contiquantumprobbessel}).

To investigate fluctuations for continuous-time QRWs, we consider both instantaneous uniform mixing property (see below) and temporal standard deviation (see Section 4). The former corresponds to a spatial fluctuation and the latter corresponds a temporal one. 

A random walk starting from $0$ on $C_N$ has the instantaneous uniform mixing property (IUMP) if there exists $t > 0$ such that $P_N (n,t) = 1/N$ for any $n=0,1, \ldots , N-1.$ To restate this definition, we introduce the total variation distance which is the notion of distance between probability distributions defined as
\begin{eqnarray}
|| P - Q || = \max_{A \subset C_N} |P(A) - Q(A)|
= {1 \over 2} \sum_{n \in C_N} |P(n) - Q(n)|
\label{eqn:variationdistance}
\end{eqnarray}
By using this, the IUMP can be rewritten as follows: if there exists $t > 0$ such that $|| P_N (\cdot,t)  - \pi_N ||=0$, where $\pi_N$ is the uniform distribution on $C_N$. As we will mention in the next section, continuous-time QRWs on $C_N$ for $N=3,4$ have the IUMP. On the other hand, Ahmadi et al. \cite{ABT} conjectured that continuous-time QRWs on $C_N$ for $N \ge 5$ do not have the IUMP. It will be stated in the last section that the continuous-time classical random walk on $C_N$ does not have the IUMP for any $N \ge 3.$ In our previous paper \cite{IKK}, we consider the temporal standard deviation for discrete-time QRWs on $C_N$, in particular, when $N$ is odd. Similarly we study it here for continuous-time case and clarify some differences between both cases. The definition of the temporal standard deviation will be given in Section 4.

\par
The structure of the present letter is as follows. Section 2 is devoted to a review on QRWs on circles. In Section 3, we consider probability distributions and time-averaged distributions for QRWs on $C_N$. Section 4 gives our new results on temporal standard deviation on continuous-time quantum case comparing with discrete-time quantum case and classical case. In the last section, we discuss a future problem on convergence.

\section{Classical Case}
First we review the classical case on $C_N$, (see Schinazi \cite{S} and references in his book, for example). In the continuous-time classical random walk, 

\begin{eqnarray*}
P_{N}(n,t)= {1 \over N} \sum_{j=0} ^{N-1} \cos (\xi_j n) e^{t(\cos \xi_j -1)},
\label{eqn:conticlassicalprob}
\end{eqnarray*}
for any $t \ge 0$ and $n=0,1, \ldots, N-1$. In this walk, the distribution for the random time between two jumps is the exponential distribution. As for the discrete-time classical random walk, 

\begin{eqnarray*}
P_{N}(n,t)= {1 \over N} \sum_{j=0} ^{N-1} \cos (\xi_j n) (\cos \xi_j)^t, 
\label{eqn:discreteclassicalprob}
\end{eqnarray*}
for any $t =0,1, \ldots,$ and $n=0,1, \ldots, N-1$. Here a time-averaged distribution in the continuous-time case is given by
\begin{eqnarray}
\bar{P}_{N}(n)=\lim_{T \rightarrow \infty} \frac{1}{T} \int_{0}^{T} P_{N}(n,t) dt ,
\label{eqn:averageconti}
\end{eqnarray}
if the right-hand side of Eq. (\ref{eqn:averageconti}) exists. As for the discrete-time case, 
\begin{eqnarray}
\bar{P}_{N}(n)=\lim_{T \rightarrow \infty} \frac{1}{T} \sum_{t=0}^{T-1} P_{N}(n,t),
\label{eqn:averagediscrete}
\end{eqnarray}
if the right-hand side of Eq. (\ref{eqn:averagediscrete}) exists. Concerning the continuous-time classical case, it is easily shown that $P_{N}(n,t) \to 1/N \> ( t \to \infty)$ for any $n=0,1, \ldots, N-1$. Then we have immediately $\bar{P}_{N}(n)= 1/N$ for any $n=0,1, \ldots, N-1$. On the other hand, in the discrete-time classical case, if $N$ is odd, then $P_{N}(n,t) \to 1/N$ as $t \to \infty$, and, if $N$ is even, then it does not converge in the limit of $t \to \infty$. However, an important property is that $\bar{P}_{N}$ is uniform distribution in both continuous- and discrete-time classical cases.

\section{Quantum Case}
\hspace*{1em}
Now we consider continuous-time QRWs. As we will see later, in contrast to classical case, $\bar{P}_{N}$ does not become uniform distribution for any $N \ge 3$ in quantum case. From Eq. (\ref{eqn:daiji}), we have
\begin{eqnarray}
P_{N}(n,t)= {1 \over N} + {2 R_N(n,t) \over N^2}, 
\label{eqn:contiquantumprob}
\end{eqnarray}
where
\begin{eqnarray}
R_N (n,t)= \sum_{0 \le j<k \le N-1} 
\cos \{ t(\cos \xi_j - \cos \xi_k) - n (\xi_j - \xi_k) \},
\label{eqn:naomi}
\end{eqnarray}
for any $t \ge 0$ and $n=0,1, \ldots, N-1$. It is easily checked that
\begin{eqnarray*}
R_N (n,t)= R_N (N-n,t),
\end{eqnarray*}
for any $t \ge 0$ and $n=1, \ldots, \bar{N}$, where $\bar{N}=[(N-1)/2]$ and $[x]$ is the smallest integer greater than $x$. For example, when $N=3$, 
\begin{eqnarray*}
R_3 (0,t) &=& 2 \cos(3t/2)+ 1, \\
R_3 (1,t) &=& R_3 (2,t) = - \cos(3t/2) - 1/2,
\end{eqnarray*}
when $N=4$,
\begin{eqnarray*}
R_4 (0,t) &=& \cos (2t) + 4 \cos (t) + 1, \\
R_4 (1,t) &=& R_4 (3,t) = - \cos (2t) - 1, \\
R_4 (2,t) &=& \cos (2t) - 4 \cos (t) + 1. \\
\end{eqnarray*}
So if $N=3$, then $R_3 (n,t) = 0$ for any $t=\pm 4\pi/9 + 4 m \pi/3 \> (m \in {\bf Z})$ and $n=0,1,2$, and if $N=4$, then $R_4 (n,t) = 0$ for any $t=\pi/2 + m \pi \> (m \in {\bf Z})$ and $n=0,1,2,3.$ Therefore continuous-time QRWs on $C_3$ and $C_4$ have IUMP. This result was given in Ref.  \cite{ABT}. On the other hand, when $N=6$,
\begin{eqnarray*}
R_6 (0,t) &=& \cos (2t) + 4 \cos(3t/2) + 4 \cos (t) + 4 \cos(t/2)+2, \\
R_6 (1,t) &=& R_6 (5,t) = - \cos (2t) - 2 \cos(3t/2) - \cos (t) + 2 \cos(t/2)-1, \\
R_6 (2,t) &=& R_6 (4,t) = \cos (2t) -2 \cos(3t/2) +  \cos (t) - 2 \cos(t/2) -1, \\
R_6 (3,t) &=& - \cos (2t) + 4 \cos(3t/2) - 4 \cos (t) - 4 \cos(t/2)+2.
\end{eqnarray*}
Direct computation implies that a continuous-time QRW on $C_6$ does not have IUMP.
\par
Now we consider the time-averaged distribution for continuous-time case. By using Eq. (\ref{eqn:contiquantumprob}), we have
\begin{eqnarray}
\bar{P}_{N}(n)= {1 \over N} + {2 R_N (n) \over N^2} ,
\label{eqn:contiquantumaverage}
\end{eqnarray}
for any $n=0,1, \ldots, N-1$, where 
\begin{eqnarray*}
R_N (n) =\left\{ 
\begin{array}{ll}
-1/2 & \qquad \hbox{if} \>\> N = \hbox{odd}, \quad \xi_{2n} \not= 0 \quad (\hbox{mod} \> 2 \pi), 
\\
-1 & \qquad \hbox{if} \>\> N = \hbox{even}, \quad \xi_{2n} \not= 0 \quad (\hbox{mod} \> 2 \pi), 
\\
 \bar{N}  & \qquad \hbox{if} \>\> \xi_{2n} = 0 \quad (\hbox{mod} \> 2 \pi). \\
\end{array} \right.
\label{eqn:rNn}
\end{eqnarray*}
This result gives
\begin{eqnarray}
\bar{P}_{N}(0)= {1 \over N} + { 2 \bar{N} \over N^2}. 
\label{eqn:pbarN0}
\end{eqnarray}
As for discrete-time quantum case given by the Hadamard transformation, Aharonov et al. \cite{AAK} and Bednarska et al. \cite{BGK} showed that $\bar{P}_{N}(n) \equiv 1/N$ if $N$ is odd or $N=4$, and $\bar{P}_{N}(n) \not\equiv 1/N $, otherwise. So in contrast to classical continuous- and discrete-time and quantum discrete-time cases, $\bar{P}_N$ in quantum continuous-time case is not uniform distribution on $C_N$ for any $N \ge 3$. In fact, when $N =$ odd (i.e., $\bar{N}=(N-1)/2$),
\begin{eqnarray*}
\bar{P}_N = \Bigl( {1 \over N}+{N-1 \over N^2}, 
\overbrace{{1 \over N} - {1 \over N^2}, \ldots , {1 \over N} - {1 \over N^2}}^{N-1} \Bigr),
\end{eqnarray*}
when $N =$ even (i.e., $\bar{N}=(N-2)/2$),
\begin{eqnarray*}
\bar{P}_N = \Bigl( {1 \over N}+{N-2 \over N^2}, 
\overbrace{{1 \over N} - {2 \over N^2}, \ldots , {1 \over N} - {2 \over N^2}}^{(N-2)/2}, {1 \over N}+{N-2 \over N^2}, 
\overbrace{{1 \over N} - {2 \over N^2}, \ldots , {1 \over N} - {2 \over N^2}}^{(N-2)/2} \Bigr).
\end{eqnarray*}
For examples, 
\begin{eqnarray*}
\bar{P}_3 = \left( {5 \over 9}, {2 \over 9}, {2 \over 9} \right),  
\qquad  
\bar{P}_4 = \left( {3 \over 8}, {1 \over 8}, {3 \over 8}, {1 \over 8} \right).
\end{eqnarray*}

\section{Temporal Standard Deviation}
We consider the following temporal standard deviation $\sigma_{N}(n)$ in the continuous-time case as in the discrete-time case:
\begin{eqnarray}
\sigma_{N}(n)=  \lim_{T \rightarrow \infty} \sqrt{
\frac{1}{T} 
\int_{0}^{T} \left (P_{N}(n,t)-\bar{P}_{N}(n) \right )^{2} dt
},
\label{eqn:sigmaconti}
\end{eqnarray}
if the right-hand side of Eq. (\ref{eqn:sigmaconti}) exists. In the discrete-time case, 
\begin{eqnarray}
\sigma_{N}(n)=  \lim_{T \rightarrow \infty} \sqrt{
\frac{1}{T} 
\sum_{t=0}^{T-1} \left (P_{N}(n,t)-\bar{P}_{N}(n) \right )^{2}
},
\label{eqn:sigmadis}
\end{eqnarray}
if the right-hand side of Eq. (\ref{eqn:sigmadis}) exists. In both continuous- and discrete-time classical cases, 
\begin{eqnarray*}
\sigma_{N}(n)=0,
\end{eqnarray*}
for $N \ge 3$ and $n=0,1, \ldots, N-1$. The reason is as follows. In the case of classical random walk starting from a site for $N=$ odd (i.e., aperiodic case), a coupling method implies that there exist $a \in (0,1)$ and $C >0$ (are independent of $n$ and $t$) such that
\begin{eqnarray*}
 |P_{N}(n,t)-\bar{P}_{N}(n)| \le C a^t,
\end{eqnarray*}
for any $n$ and $t$ (see page 63 of Schinazi \cite{S}, for example). Therefore we obtain
\begin{eqnarray*}
\frac{1}{T} \sum_{t=0}^{T-1} \left (P_{N}(n,t)-\bar{P}_{N}(n) \right )^{2}
\le {C^2 \over T} {1-a^{2T} \over 1-a^2}.
\end{eqnarray*}
The above inequality implies that $\sigma_{N}(n)=0 \> (n=0,1, \ldots, N-1)$. As for $N=$ even (i.e., periodic) case, we have the same conclusion $\sigma_{N}(n)=0$ for any $n$ by using a little modified argument. The same conclusion can be extended to continuous-time classical case in a standard fashion.

For discrete-time quantum case, our previous paper \cite{IKK} implies
\begin{eqnarray}
\sigma_{N}^{2}(n)=\frac{1}{N^{4}} 
\left[
2 
\left\{ S_{+}^{2}(n)+S_{-}^{2}(n) \right\}
+11S_{0}^{2}+10 S_{0}S_{1}+3S_{1}^{2}-S_{2}(n)
\right]
-\frac{2}{N^{3}},
\label{eqn:sigmaF}
\end{eqnarray}
for any $n=0,1, \ldots, N-1$, where $N (\ge 3)$ is odd and
\begin{eqnarray*}
S_{0} &=& \sum_{j=0}^{N-1} \frac{1}{3+\cos \theta_{j} }, \qquad 
S_{1} \>\> = \>\> 
\sum_{j=0}^{N-1} \frac{\cos \theta_{j}}{3+\cos \theta_{j}}, \\
S_{+}(n) &=& \sum_{j=0}^{N-1} \frac{  \cos \left(  (n-1) \theta_{j} \right)+\cos \left(n \theta_{j} \right) }
                  {3+\cos \theta_{j}}, \\
S_{-}(n) &=& \sum_{j=0}^{N-1} \frac{  \cos \left(  (n-1)\theta_{j} \right)-\cos \left( n \theta_{j}  \right)}
                  {3+\cos  \theta_{j} }, \\
S_{2}(n) &=& \sum_{j=1}^{N-1} \frac{7+\cos   (2 \theta_{j}) 
                                    +8\cos     \theta_{j}\cos^{2} 
                                            \left( \left(n-1/2 \right)
                                             \theta_{j}  \right) }
                                   {(3+\cos \theta_{j})^{2}},
\label{functions}
\end{eqnarray*}
with $\theta_{j} = \xi_{2j} = 4 \pi j/N$. For example,
\begin{eqnarray}
\sigma_3(0)=\sigma_3(1)={2 \sqrt{46} \over 45}, \quad \sigma_3(2)={2 \over 9}.
\label{eqn:aki}
\end{eqnarray}
Furthermore we \cite{IKK} obtained 
\begin{eqnarray}
\sigma_{N}(0) = \frac{\sqrt{13-8\sqrt{2}}}{N}+ o \left( {1 \over N} \right)
\label{eqn:daijiyo}	
\end{eqnarray}
as $N \to \infty$. The above result implies that the temporal fluctuation $\sigma_{N}(0)$ in discrete-time case decays in the form $1/N$ as $N$ increases.

Now we consider continuous-time quantum case. For example, direct computation gives 
\begin{eqnarray*}
&& 
\sigma_3(0)={2 \sqrt{2} \over 9}, \quad \sigma_3 (1)= \sigma_3(2)={\sqrt{2} \over 9}, \\
&&
\sigma_4(0)=\sigma_4(2)={\sqrt{34} \over 16}, \quad \sigma_4(1)=\sigma_4(3)={\sqrt{2} \over 16},
\\
&&
\sigma_5(0)={4 \sqrt{3} \over 25}, \quad \sigma_5(1)=\sigma_5(2)=\sigma_5(3)=\sigma_5(4)={2 \sqrt{2} \over 25}, \\
&&
\sigma_6(0)=\sigma_6(3)={7 \sqrt{2} \over 36}, \quad \sigma_6(1)=\sigma_6(2)=\sigma_5(4)=\sigma_5(5)={\sqrt{5} \over 18}.
\end{eqnarray*}

From now on we focus on general $N=$ odd case to obtain results corresponding to discrete-time case (i.e., Eqs. (\ref{eqn:sigmaF}) and (\ref{eqn:daijiyo})). The definition of $\sigma_{N} (n)$ implies 
\begin{eqnarray*}
\sigma_{N}^2(n) 
&=& 
{4 \over N^4} \times \lim_{T \rightarrow \infty} 
\frac{1}{T} 
\int_{0}^{T} \left( R_{N}(n,t) - R_{N}(n) \right)^2 dt \\
&=& 
{4 \over N^4} \times \left\{ 
\lim_{T \rightarrow \infty} 
\frac{1}{T} 
\int_{0}^{T} R_{N}(n,t)^2 dt - R_{N}(n)^2 \right\} \\
%\label{eqn:sigmaconti_2}
\end{eqnarray*}
For $N=$ odd case, we have
\begin{eqnarray*}
R_N (n) =\left\{ 
\begin{array}{ll}
(N-1)/2 & \qquad n = 0, 
\\
-1/2 & \qquad n = 1, 2, \ldots, N-1.
\end{array} \right.
\label{eqn:rNnodd}
\end{eqnarray*}
From Eq. (\ref{eqn:naomi}), we have 
\begin{eqnarray*}
&&
\lim_{T \rightarrow \infty} 
\frac{1}{T} 
\int_{0}^{T} 
R_N ^{2}(n,t)
dt 
\\
&=&
\lim_{T \rightarrow \infty}
\frac{1}{T}
\int_{0}^{T}
\left[
\sum_{0 \le j_{1}<k_{1} \le N-1}
\cos 
\{ t(\cos \xi_{j_{1}} - \cos \xi_{k_{1}}) - n (\xi_{j_{1}} - \xi_{k_{1}})\}
\right] \nonumber \\
&&\hspace{18mm}
\times
\left[
\sum_{0 \le j_{2}<k_{2} \le N-1} 
\cos 
\{ t(\cos \xi_{j_{2}} - \cos \xi_{k_{2}}) - n (\xi_{j_{2}} - \xi_{k_{2}}) \}
\right]
dt \nonumber \\
&=&
\lim_{T \rightarrow \infty} 
\frac{1}{2T} 
\int_{0}^{T}
dt
\sum_{0 \le j_{1}<k_{1} \le N-1} 
\sum_{0 \le j_{2}<k_{2} \le N-1} 
\nonumber \\
&&
\hspace{-9mm}
\left[
\cos \{ t((\cos \xi_{j_{1}} - \cos \xi_{k_{1}})-(\cos \xi_{j_{2}} - \cos \xi_{k_{2}})) - n ((\xi_{j_{1}} - \xi_{k_{1}}) -(\xi_{j_{2}} - \xi_{k_{2}})) \}
\right.
\nonumber \\
&&
\hspace{-7mm}
\left.
+
\cos \{ 
t((\cos \xi_{j_{1}} - \cos \xi_{k_{1}})+(\cos \xi_{j_{2}} - \cos \xi_{k_{2}})) - n ((\xi_{j_{1}} - \xi_{k_{1}}) +(\xi_{j_{2}} - \xi_{k_{2}})) \}
\right]
\nonumber \\
&&
\label{eqn:sigmaconti_2_1_3}
\end{eqnarray*}
Remark that each $(N^2-N)^2$ term becomes zero except the following two cases:
\begin{eqnarray*}
\cos \xi_{j_{1}} - \cos \xi_{k_{1}}-\cos \xi_{j_{2}} + \cos \xi_{k_{2}}=0, 
\\
\cos \xi_{j_{1}} - \cos \xi_{k_{1}}+\cos \xi_{j_{2}} - \cos \xi_{k_{2}}=0.
\end{eqnarray*}
Therefore some complicated computations imply
\begin{eqnarray*}
\lim_{T \rightarrow \infty} 
\frac{1}{T} 
\int_{0}^{T} 
R_N^{2} (0,t)
dt &=&
{5 \over 8}(N-1)^2 
+{1 \over 8} (N-1)(5N-13)
\nonumber \\
&=&{1 \over 4}(N-1)(5N-9)
\label{eqn:konishikun}
\end{eqnarray*}
and
\begin{eqnarray*}
\lim_{T \rightarrow \infty} 
\frac{1}{T} 
\int_{0}^{T} 
R_N^{2} (n,t)
dt &=&
{{1} \over 8} 
\left( 2 N^2 -4 N +6 -{{1} \over {\cos ^2{{{n \pi} \over N}}}} \right)
\\
&&
\qquad \qquad \qquad \qquad 
+{{1} \over 8}
\left( -6 N +12 +{{1} \over {\cos ^2{{{n \pi} \over N}}}} \right)
\nonumber \\
&=&{{1} \over 4}(N^2 -5 N +9)
\label{eqn:kasaharakun}
\end{eqnarray*}
for any $n=1,2, \ldots , N-1.$ Then we have the following main result: when $N (\ge 3)$ is odd, 
\begin{eqnarray}
\sigma_N (0) = {2\sqrt{N^2 -3 N + 2} \over N^2}, \quad 
\sigma_N (n) = {\sqrt{N^2 -5 N +8} \over N^2} \quad (n=1, \ldots, N-1).
\label{eqn:suki}
\end{eqnarray}
In particular, $\sigma_N (0) > \sigma_N (n)$ for any $n=1, \ldots, N-1.$ We should remark that there is a difference between continuous-time case and discrete-time one, since, for example, Eq. (\ref{eqn:aki}) gives $\sigma_3 (0) = \sigma_3 (1) > \sigma_3 (2)$ in discrete-time case. Furthermore Eq. (\ref{eqn:suki}) implies 
\begin{eqnarray} 
\sigma_{N}(0) = \frac{2}{N} + o \left( {1 \over N} \right), \quad 
\sigma_{N}(n) = \frac{1}{N} + o \left( {1 \over N} \right) 
\quad (n=1, \ldots, N-1)
\label{eqn:sukisuki}	
\end{eqnarray}
as $N \to \infty$. The above result implies that the temporal fluctuation $\sigma_{N}(0)$ in continuous-time case decays in the form $1/N$ as $N$ increases as in the discrete-time case. However, from $\sqrt{13 - 8 \sqrt{2}} < 2$ (see Eqs. (\ref{eqn:daijiyo}) and (\ref{eqn:sukisuki})), it is obtained that a temporal fluctuation of continuous-time case at site 0 is greater than that of discrete-time case for large $N$.

\section{Future Problem}
Finally we consider a future problem on the total variation convergence and weak convergence. We should remark that the total variation convergence is stronger than weak convergence. Parity problem vanishes in continuous-time case, so we focus on continuous-time case even in the classical case. Now we introduce
\begin{eqnarray*}
d_N (t) = || P_N (\cdot,t)  - \pi_N|| = {1 \over 2} \sum_{n \in C_N} |P_N (n,t) - \pi_N (n)|,
\label{eqn:variationdistancednt}
\end{eqnarray*}
where $\pi_N$ is the uniform distribution on $C_N$. In the classical continuous-time case, $d_N(t)$ decreases as $t$ increases, and
\begin{eqnarray*}
\lim_{t \to \infty} d_N (t) = 0.
\label{eqn:variationdistancezero}
\end{eqnarray*}
So the continuous-time classical random walk on $C_N$ does not have the IUMP for any $N \ge 3.$ Moreover, it is known that
\begin{eqnarray*}
\lim_{N \to \infty} d_N (t N^2) 
= d_{\infty} (t) \equiv {1 \over 2} \int_0 ^1 |f_t (x) - 1| dx,
\label{eqn:variationdistancebm}
\end{eqnarray*}
where $f_t$ is the density of $\sqrt{t}Z$ mod 1 and $Z$ has the standard normal distribution (see Chapters 2 and 5 in \cite{AF}). This fact corresponds to weak convergence of classical random walk to Brownian motion on a torus ${\bf T}$, i.e., ${\bf R}$ mod 1. In the continuous-time quantum case, an interesting future problem is to obtain a limit theorem corresponding to the above classical case.

\section*{Acknowledgements}
This work is partially financed by the Grant-in-Aid for Scientific Research (B) (No.12440024) of Japan Society of the Promotion of Science. 

\appendix % Reset the environments to Appendix style

%%%%%%%%%%%%%%%%%%%%%%%%%%%%%%
% For BiBTeX users, just uncomment the following two lines
%\bibliographystyle{unsrt}
%\bibliography{mybibfile}

\begin{small}

\par
\vskip 1.0cm
Graduate School of Engineering 

University of Hyogo

2167 Syosha, Himeji, 671-2203, Japan

inui@eng.u-hyogo.ac.jp

\par
\vskip 1.0cm
Department of Applied Mathematics

Faculty of Engineering

Yokohama National University

Hodogaya-ku, Yokohama 240-8501, Japan

KSHR@lam.osu.sci.ynu.ac.jp, norio@mathlab.sci.ynu.ac.jp, 

\par
\vskip 1.0cm
Department of Mechanical and Intelligent Engineering

Himeji Institute of Technology

2167 Syosha, Himeji, 671-2203, Japan

\end{small}

\end{document}